\documentclass{llncs}

\usepackage[pdftex]{hyperref}
\usepackage{xspace}
\newcommand{\kblog}{kblog\xspace}

\begin{document}
\title{Three Steps to Heaven: Semantic Publishing in a Real World Workflow}
\author{Phillip Lord\inst{1}, Simon Cockell\inst{2} and  Robert Stevens\inst{3}}

\institute{School of Computing Science, Newcastle University,
Newcastle-upon-Tyne, UK \and 
Bioinformatics Support Unit, Newcastle University,
Newcastle-upon-Tyne, UK \and
School of Computer Science, University of Manchester, UK
\email{phillip.lord@newcastle.ac.uk}}

\maketitle

\begin{abstract}
  Semantic publishing offers the promise of computable papers, enriched
  visualisation and a realisation of the linked data ideal. In reality,
  however, the publication process contrives to prevent richer semantics while
  culminating in a `lumpen' PDF. In this paper, we discuss a web-first
  approach to publication, and describe a three-tiered approach which
  integrates with the existing authoring tooling. Critically, although it adds
  limited semantics, it does provide value to all the participants in the
  process: the author, the reader and the machine.
  
  \textbf{License:} This work is licensed under a Creative Commons Attribution
  3.0 Unported License. http://creativecommons.org/licenses/by/3.0/. It is
  also available at 
  \href{http://www.russet.org.uk/blog/2012/04/three-steps-to-heaven/}
  {http://www.russet.org.uk/blog/2012/04/three-steps-to-heaven/}
\end{abstract}

\section{Introduction}
\label{sec:introduction}

The publishing of both data and narratives on those data are changing
radically. Linked Open Data and related semantic technologies allow for
semantic publishing of data. We still need, however, to publish the narratives
on that data and that style of publishing is in the process of change; one of
those changes is the incorporation of
semantics~\cite{shadbolt2006semantic,shotton2009semantic,shotton2009adventures}.
The idea of semantic publishing is an attractive one for those who wish to
consume papers electronically; it should enhance the richness of the
computational component of papers~\cite{shotton2009semantic}. It promises a
realisation of the vision of a next generation of the web, with papers
becoming a critical part of a linked data
environment~\cite{shadbolt2006semantic,bizer2009linked}, where the results and
naratives become one.

The reality, however, is somewhat different. There are significant barriers to
the acceptance of semantic publishing as a standard mechanism for academic
publishing. The web was invented around 1990 as a light-weight mechanism for
publication of documents. It has subsequently had a massive impact on society
in general. It has, however, barely touched most scientific publishing; while
most journals have a website, the publication process still revolves around
the generation of papers, moving from Microsoft Word or \LaTeX~\cite{latex}, through to
a final PDF which looks, feels and is something designed to be printed onto
paper\footnote{This includes conferences dedicated to the web and the use of
  web technologies.}. Adding semantics into this environment is difficult or
impossible; the content of the PDF has to be exposed and semantic content
retro-fitted or, in all likelihood, a complex process of author and publisher
interaction has to be devised and followed. If semantic data publishing and
semantic publishing of academic narratives are to work together, then academic
publishing needs to change.

In this paper, we describe our attempts to take a commodity publication
environment, and modify it to bring in some of the formality required from
academic publishing. We illustrate this with three exemplars---different kinds
of knowledge that we wish to enhance. In the process, we add a small amount of
semantics to the finished articles. Our key constraint is the desire to add
value for all the human participants. Both authors and readers should see and
recognise additional value, with the semantics a useful or necessary byproduct
of the process, rather than the primary motivation. We characterise this
process as our ``three steps to heaven'', namely:

\begin{itemize}
\item make life better for the machine to
\item make life better for the author to
\item make life better for the reader
\end{itemize}

While requiring additional value for all of these participants is hard, and
places significant limitations on the level of semantics that can be achieved,
we believe that it does increase the likelihood that content will be generated in
the first place, and represents an attempt to enable semantic publishing in a
real-world workflow.

\section{Knowledgeblog}
\label{sec:knowledgeblog}

The \href{http://knowledgeblog.org}{knowledgeblog} project stemmed from the
desire for a book describing the many aspects of ontology development, from
the underlying formal semantics, to the practical technology layer and,
finally, through to the knowledge domain~\cite{stlr_kblog_2011}. However, we
have found the traditional book publishing process frustrating and
unrewarding. While scientific authoring is difficult in its own right, our own
experience suggests that the \emph{publishing} process is extremely hard-work.
This is particularly so for multi-author collected works which are often
harder for the editor than writing a book ``solo''. Finally, the expense and
hard copy nature of academic books means that, again in our experience, few
people read them.

This contrasts starkly with the web-first publication process that has become
known as blogging. With any of a number of ready made platforms, it is
possible for authors with little or no technical skill, to publish content to
the web with ease. For knowledgeblog (``kblog''), we have taken one blogging
engine, WordPress~\cite{wordpress}, running on low-end hardware, and used it
to develop a multi-author resource describing the use of ontologies in the
life sciences (our main field of expertise). There are also kblogs on
bioinformatics\footnote{\url{http://bioinformatics.knowledgeblog.org}} and the
Taverna workflow
environment\footnote{\url{http://taverna.knowledgeblog.org}}~\cite{Hull:2006:Nucleic-Acids-Res:16845108}.
We have previously described how we addressed some of the social aspects,
including attribution, reviewing and immutablity of
articles\cite{stlr_kblog_2011}.  

As well as delivering content, we are also using this framework to investigate
\emph{semantic academic publishing}, investigating how we can enhance the
machine interpretability of the final paper, while living within the key
constraint of making life (slightly) better for machine, author and reader
without adding complexity for the human participants.

Scientific authors are relatively conservative. Most of them have
well-established toolsets and workflows which they are relatively unwilling to
change. For instance, within the kblog project, we have used workshops to
start the process of content generation. For our initial meeting, we gave
little guidance on authoring process to authors, as a result of which most
attempted to use WordPress directly for authoring. The WordPress editing
environment is, however, web-based, and was originally designed for editing
short, non-technical articles. It appeared to not work well for most
scientists.

The requirements that authors have for such `scientific' articles are
manifold. Many wish to be able to author while offline (particularly on trains
or planes). Almost all scientific papers are multi-author, and some degree of
collaboration is required. Many scientists in the life sciences wish to author
in Word because grant bodies and journals often produce templates as Word
documents. Many wish to use \LaTeX{}, because its idiomatic approach to
programming documents is unreplicable with anything else. Fortunately, it is
possible to induce WordPress to accept content from many different authoring
tools, including Word and \LaTeX{}\cite{stlr_kblog_2011}.

As a result, during the kblog project, we have seem many different
workflows in use, often highly idiosyncratic in nature. These include:
\begin{description}
\item[Word/Email:] Many authors write using MS~Word and collaborate by emailing
  files around. This method has a low barrier to entry, but requires
  significant social processes to prevent conflicting versions, particularly
  as the number of authors increases.
\item[Word/Dropbox:] For the \href{http://taverna.knowledgeblog.org}{taverna}
  kblog, authors wrote in Word and collaborated with
  Dropbox.\footnote{\url{http://www.dropbox.com}} This method works reasonably
  well where many authors are involved; Dropbox detects conflicts, although
  it cannot prevent or merge them.
\item[Asciidoc/Dropbox:] Used by the authors of this paper.
  Asciidoc\footnote{\url{http://www.methods.co.nz/asciidoc/}} is relatively simple,
  somewhat programmable and accessible. Unlike \LaTeX{} which can be induced to
  produce HTML with effort, asciidoc is designed to do so.
\end{description}

Of these three approaches probably the Word/Dropbox combination is the
the most generally used.

From the readers perspective, a decision that we have made within
knowledgeblog is to be ``HTML-first''. The initial reasons for this were
entirely practical; supporting multiple toolsets is hard, particularly if any
degree of consistency is to be maintained; the generation of the HTML is at
least partly controlled by the middleware -- WordPress in \kblog's case. As
well as enabling consistency of presentation, it also, potentially, allows us
to add additional knowledge; it makes semantic publication a possibility.
However, we are aware that knowledgeblog currently scores rather badly on what
we describe as the ``bath-tub test''; while exporting to PDF or printing out
is possible, the presentation is not as ``neat'' as would be ideal. In this
regard (and we hope only in this regard), the knowledgeblog experience is
limited. However, increasingly, readers are happy and capable of interacting
with material on the web, without print outs.

From this background and aim, we have drawn the following requirements:
\begin{enumerate}
\item The author can, as much as possible, remain within familiar authoring
  environments;
\item The representation of the published work should remain extensible to,
  for instance, semantic enhancements;
\item The author and reader should be able to have the amount of ``formal''
  academic publishing they need;
\item Support for semantic publishing should be gradual and offer advantages
  for author and reader at all stages.
\end{enumerate}

We describe how we have achieved this with three exemplars, two of which are
relatively general in use, and one more specific to biology. In each case, we
have taken a slightly different approach, but have fulfilled our primary aim of
making life better for machine, author and reader.

\section{Representing Mathematics}
\label{sec:repr-math}

The representation of mathematics is a common need in academic literature.
Mathematical notation has grown from a requirement for a syntax which is
highly expressive and relatively easy to write. It presents specific
challenges because of its complexity, the difficulty of authoring and the
difficulty of rendering, away from the chalk board that is its natural home.

Support for mathematics has had a significant impact on academic publishing.
It was, for example, the original motivation behind the development of
\TeX~\cite{tex}, and it still one of the main reasons why authors wish to use
it or its derivatives. This is to such an extent that much mathematics
rendering on the web is driven by a \TeX{} engine somewhere in the process. So
MediaWiki (and therefore Wikipedia), Drupal and, of course, WordPress follow
this route. The latter provides plugin support for \TeX{} markup using the
\texttt{wp-latex} plugin \cite{wp-latex}. Within kblog, we have developed a
new plugin called \texttt{mathjax-latex} \cite{mathjax-latex}. From the \kblog
author's perspective these two offer a similar interface -- differences are,
therefore, described later.

Authors write their mathematics directly as \TeX{} using one of the four
markup syntaxes. The most explicit (and therefore least likely to happen
accidentally) is through the use of
``shortcodes''.\footnote{\url{http://codex.wordpress.org/Shortcode}} These are
a HTML-like markup originating from some forum/bulletin board systems. In this
form an equation would be entered as \verb#[latex]e=mc^2[/latex]#, which would
be rendered as ``$e=mc^2$''. It is also possible to use three other syntaxes which
are closer to math-mode in \TeX: \verb#$$e=mc^2$$#, \verb#$latex e=mc^2$#, or
\verb#\[e=mc^2\]#.

From the authorial perspective, we have added significant value, as it is
possible to use a variety of syntaxes, which are independent of the authoring
engine. For example, a \TeX-loving mathematician working with a Word-using
biologist can still set their equations using \TeX{} syntax; although Word
will not render these at authoring time but, in practice, this causes few
problems for such authors, who are experienced at reading \TeX{}. Within an
\LaTeX{} workflow equations will be renderable both locally with source compiled
to PDF, and published to WordPress.

There is also a W3C recommendation, MathML for the representation and
presentation of mathematics. The kblog environment also supports this. In this
case, the equivalent source appears as follows:

\begin{verbatim}
<math>
    <mrow>
        <mi>E</mi>
        <mo>=</mo>
        <mrow>
            <mi>m</mi>
            <msup>
                <mi>c</mi>
                <mn>2</mn>
            </msup>
        </mrow>
    </mrow>
</math>
\end{verbatim}

One problem with the MathML representation is obvious: it is very long-winded.
A second issue, however, is that it is hard to integrate with existing
workflows; most of the publication workflows we have seen in use will on
recognising an angle bracket turn it into the equivalent HTML entity. For some
workflows (\LaTeX, asciidoc) it is \emph{possible}, although not easy, to
prevent this within the native syntax.

It is also possible to convert from Word's native OMML (``equation
editor'') XML representation to MathML, although this does not integrate with
Word's native blog publication workflow. Ironically, it is because MathML
shares an XML based syntax with the final presentation format (HTML) that the
problem arises. The shortcode syntax, for example, passes straight-through
most of the publication frameworks to be consumed by the middleware. From a
pragmatic point of view, therefore, supporting shortcodes and \TeX-like
syntaxes has considerable advantages.

For the reader, the use of \texttt{mathjax-latex} has significant advantages.
The default mechanism within WordPress uses a math-mode like syntax
\verb#$latex e=mc^2$#. This is rendered using a \TeX{} engine into an image
which is then incorporated and linked using normal HTML capabilities. This
representation is opaque and non-semantic; it has significant limitations
for the reader. The images are not scalable -- zooming in cases severe
pixalation; the background to the mathematics is coloured inside the image, so
does not necessarily reflect the local style.

Kblog, however, uses the MathJax library\cite{mathjax}; this has a number of
significant advantages for the reader. First, where the browser supports them,
MathJax uses webfonts to render the images; these are scalable, attractive and
standardized. Where they are not available, MathJax can fall-back to bitmapped
fonts. The reader can also access additional functionality: clicking on an
equation will raise a zoomed in popup; while the context menu allows access to
a textual representation either as \TeX{} or MathML irrespective of the form
that the author used. This can be cut-and-paste for further use. Kblog uses
the MathJax library\cite{mathjax} to render the underlying \TeX{} directly on
the client.

Our use of MathJax provides no significant disadvantages to the middleware
layers. It is implemented in JavaScript and runs in most environments.
Although, the library is fairly large ($>$100Mb), but is available on a CDN so
need not stress server storage space. Most of this space comes from the
bit-mapped fonts which are only downloaded on-demand, so should not stress web
clients either. It also obviates the need for a \TeX{} installation which
\texttt{wp-latex} may require (although this plugin can use an external server
also).

At face value, \texttt{mathjax-latex} necessarily adds very little semantics to the
maths embedded within documents. The maths could be represented as
\verb#$$E=mc^2$$#, \verb#\(E=mc^2\)]# or

\begin{verbatim}
<math> <mrow> <mi>E</mi> <mo>=</mo> <mrow> <mi>m</mi>
<msup> <mi>c</mi><mn>2</mn> </msup>
</mrow> </mrow> </math>
\end{verbatim}

So, we have a heterogenous representation for identical knowledge. However, in
practice, the situation is much better than this. The author of the work
created these equations and has then read them, transformed by MathJax into a
rendered form. If MathJax has failed to translate them correctly, in line with
the author's intention, or if it has had some implications for the text in
addition to setting the intended equations (if the \TeX{} style markup appears
accidentally elsewhere in the document), the author is likely to have seen
this and fixed the problem. Someone wishing, for example, to extract all the
mathematics as MathML from these documents computationally, therefore, knows:

\begin{itemize}
\item that the document contains maths as it imports MathJax
\item that MathJax is capable of identifying this maths correctly
\item that equations can be transformed to MathML using MathJax\footnote{This
    is assuming MathJax works correctly in general. The authors and readers
    are checking the rendered representation. It is possible that an equation
    would render correctly on screen, but be rendered to MathML inaccurately}.
\end{itemize}

So, while our publication environment does not result directly in lower level
of semantic heterogeneity, it does provide the data and the tools to enable
the computational agent to make this transformation. While this is imperfect,
it should help a bit. In short, we provide a practical mechanism to identify text containing
mathematics and a mechanism to transform this to a single, standardised
representation.

\section{Representing References}
\label{sec:repr-refer}

Unlike mathematics, there is no standard mechanism for reference and in-text
citation, but there are a large number of tools for authors such as BibTeX,
Mendeley~\cite{Mendeley} or EndNote. As a result of this, the integration with
existing toolsets is of primary importance, while the representation of the
in-text citations is not, as it should be handled by the tool layer anyway.

Within kblog, we have developed a plugin called
kcite.\footnote{\url{http://wordpress.org/extend/plugins/kcite/}} For the
author, citations are inserted using the syntax:\\
\verb#[cite]10.1371/journal.pone.0012258[/cite]#.\\ The identifier used here
is a DOI, or digital object identifier and, is widely used within the
publishing and library industry. Currently, kcite supports DOIs minted by
either CrossRef\footnote{\url{http://wordpress.org/extend/plugins/kcite/}} or
DataCite\footnote{\url{http://www.datacite.org/}} (in practice, this means
that we support the majority of DOIs). We also support identifiers from
PubMed\footnote{\url{http://www.ncbi.nlm.nih.gov/pubmed/}} which covers most
biomedical publications and arXiv,\footnote{\url{http://arxiv.org/}} the
physics (and other domains!) preprints archive, and we now have a system to
support arbitrary URLs. Currently, authors are required to select the
identifier where it is not a DOI.

We have picked this ``shortcode'' format for similar reasons as described
for maths; it is relatively unambiguous, it is not XML based, so passes
through the HTML generation layer of most authoring tools unchanged and is
explicitly supported in WordPress, bypassing the need for regular expressions
and later parsing. It would, however, be a little unwieldy from the
perspective of the author. In practice, however, it is relatively easy to
integrate this with many reference managers. For example, tools such as
Zotero~\cite{zotero} and Mendeley use the Citation Style
Language, and so can output kcite compliant citations with the following
slightly elided code:

\begin{verbatim}
 <citation>
    <layout prefix="[cite]" suffix="[/cite]" 
         delimiter="[/cite] [cite]">
      <text variable="DOI"/>
    </layout>
  </citation>
\end{verbatim}

We do not yet support \LaTeX/BibTeX citations, although we see no reason why
a similar style file should not be supported. We do, however, support
BibTeX-formatted files: the first author's preferred editing/citation
environment is based around these with Emacs, RefTeX, and asciidoc. While this
is undoubtedly a rather niche authoring environment, the (slightly elided)
code for supporting this demonstrates the relative ease with which tool chains
can be induced to support kcite:

\begin{verbatim}
(defadvice reftex-format-citation (around phil-asciidoc-around activate)
  (if phil-reftex-citation-override
      (setq ad-return-value (phil-reftex-format-citation entry format))
    ad-do-it))

(defun phil-reftex-format-citation( entry format )
  (let ((doi (reftex-get-bib-field "doi" entry)))
    (format "pass:[[cite source='doi'\\]%s[/cite\\]]" doi)))
\end{verbatim}

The key decision with kcite from the authorial perspective is to ignore the
reference list itself and focus only on in-text citations, using public
identifiers to references. This simplifies the tool integration process
enormously, as this is the only data that needs to pass from the author's
bibliographic database onward. The key advantage for authors here is two-fold:
they are not required to populate their reference metadata for themselves,
and this metadata will update if it changes. Secondly, the identifiers are
checked; if they are wrong, the authors will see this straightforwardly as the
entire reference will be wrong. Adding DOIs or other identifiers moves from
becoming a burden for the author to becoming a specific advantage.

While supporting multiple forms of reference identifier (CrossRef DOI,
DataCite DOI, arXiv and PubMed ID) provides a clear advantage to the author,
it comes at considerable cost. While it is possible to get metadata about
papers from all of these sources, there is little commonality between them.
Moreover, resolving this metadata requires one outgoing HTTP
request\footnote{In practice, it is often more; DOI requests, for instance,
  use \texttt{303} redirects.} per reference, which browser security might or
might not allow.

So, while the presentation of mathematics is performed largely on the client,
for reference lists the kcite plugin performs metadata resolution and data
integration on the server. A caching functionality is provided, storing this
metadata in the WordPress database. The bibliographic metadata is finally
transferred to the client encoded as JSON, using asynchronous call-backs to
the server.

Finally, this JSON is rendered using the citeproc-js library on the client. In
our experience, this performs well, adding to the readers' experience; in-text
citations are initially shown as hyperlinks; rendering is rapid, even on aging
hardware, and finally in-text citations are linked both to the bibliography
and directly through to the external source. Currently, the format of the
reference list is fixed, however, citeproc-js is a generalised reference
processor, driven using CSL\footnote{\url{http://citationstyles.org/}}. This
makes it straight-forward to change citation format, at the option of the
reader, rather than the author or publisher. Both the in-text citation and
bibliography support outgoing links direct to the underlying
resources\footnote{Where the identifier allows -- PubMed IDs redirect to
  PubMed.}. As these links have been used to gather metadata, they are likely
to be correct. While these advantages are relatively small currently, we
believe that the use of JavaScript rendering over a linked references can be
used to add further reader value in future.

For the computational agent wishing to consume bibliographic information, we
have added significant value compared to the pre-formatted HTML reference
list. First, all the information required to render the citation is present in
the in-text citation next to the text that the authors intended. A
computational agent can, therefore, ignore the bibliography list itself
entirely. These primary identifiers are, again, likely to be correct because
the authors now need them to be correct for their own benefit.

Should the computational agent wish, the (denormalised) bibliographic data
used to render the bibliography is actually available, present in the underlying
HTML as a JSON string. This is represented in a homogeneous format, although,
of course, represents our (kcite's) interpretation of the primary data.

A final, and subtle, advantage of kcite is that the authors can only use
public metadata, and not their own. If they use the correct primary
identifier, and still get an incorrect reference, it follows that the public
metadata must be incorrect\footnote{Or, we acknowledge, that kcite is
  broken!}. Authors and readers therefore must ask the metadata providers to
fix their metadata to the benefit of all. This form of data linking,
therefore, can even help those who are not using it.

\subsection{Microarray Data}
\label{sec:microarray-data}

Many publications require that papers discussing microarray experiments lodge
their data in a publically available resource such as
ArrayExpress~\cite{Brazma01012003}. Authors do this placing an ArrayExpress
identifier which has the form \verb#E-MEXP-1551#. Currently, adding this
identifier to a publication, as with adding the raw data to the repository is
no direct advantage to the author, other than fulfilment of the publication
requirement. Similarly, there is no existing support within most authoring
environments for adding this form of reference.

For the knowledgeblog-arrayexpress
plugin,\footnote{\url{http://knowledgeblog.org/knowledgeblog-arrayexpress}}
therefore, we have again used a shortcode representation, but allowed the
author to automatically fill
metadata, direct from ArrayExpress. So a tag such as:\\
\verb#[aexp id="E-MEXP-1551"]species[/aexp]# \\
will be replaced with \textit{Saccharomyces cerevisiae}, while:\\
\verb#[aexp id="E-MEXP-1551"]releasedate[/aexp]#\\ will be replaced by
``2010-02-24''. While the advantage here is small, it is significant.
Hyperlinks to ArrayExpress are automatic, authors no longer need to look up
detailed metadata. For metadata which authors are likely to know anyway (such
as Species), the automatic lookup operates as a check that their ArrayExpress
ID is correct. As with references(see Section~\ref{sec:references}), the use of an
identifier becomes an advantage rather than a burden to the authors.

Currently, for the reader there is less significant advantage at the moment.
While there is some value to the author of the added correctness stemming from
the ArrayExpress identifier. However, knowledgeblog-arrayexpress is currently
under-developed, and the added semantics that is now present could be used
more extensively. The unambiguous knowledge
that:\\\verb#[aexp id="E-MEXP-1551"]species[/aexp]#\\ represents a species
would allow us, for example, to link to the NCBI taxonomy
database.\footnote{\url{http://www.ncbi.nlm.nih.gov/Taxonomy/}}

Likewise, advantage for the computational agent from
knowledgeblog\--array\-express is currently limited; the identifiers are clearly
marked up, and as the authors now care about them, they are likely to be correct.
Again, however, knowledgeblog\--array\-express is currently under developed for the computational agent. The
knowledge that is extracted from ArrayExpress could be presented within the
HTML generated by knowledgeblog\--array\-express, whether or not it is displayed
to the reader for, essentially no cost. By having an underlying shortcode
representation, if we choose to add this functionality to
knowledgeblog\--array\-express, any posts written using it would automatically
update their HTML. For the text-mining bioinformatician, even the ability to
unambiguously determine that a paper described or used a data set relating to
a specific species using standardised nomenclature\footnote{the standard
  nomenclature was only invented in 1753 and is still not used universally.}
would be a considerable boon.

\section{Discussion}
\label{disc}

Our approach to semantic enrichment of articles is a measured and evolutionary
approach. We are investigating how we can increase the amount of knowledge in
academic articles presented in a computationally accessible form. However, we
are doing so in an environment which does not require all the different
aspects of authoring and publishing to be over-turned. More over, we have
followed a strong principle of semantic enhancement which offers advantages to
both reader and author immediately. So, adding references as a DOI, or other
identifier, `automagically' produces an in text citation and a nicely
formatted reference list: that the reference list is no longer present in the
article, but is a visualisation over linked data; that the article itself has
become a first class citizen of this linked data environment is a happy
by-product.

This approach, however, also has disadvantages. There are a number of semantic
enhancements which we could make straight-forwardly to the knowledgeblog
environment that we have not; the principles that we have adopted requires
significant compromise. We offer here two examples.

First, there has been significant work by others on CiTO~\cite{cito} -- an
ontology which helps to describe the relationship between the citations and a
paper. Kcite lays the ground-work for an easy and straight-forward addition of
CiTO tags surrounding each in-text citation. Doing so, would enable increased
machine understandability of a reference list. Potentially, we could use this
to the advantage to the reader also: we could distinguish between reviews and
primary research papers; highlight the authors' previous work; emphasise older
papers which are being refuted. However, to do this requires additional
semantics from the author. Although these CiTO semantic enhancements would be
easy to insert directly using the shortcode syntax, most authors will want to
use their existing reference manager which will not support this form of
semantics; even if it does, the author themselves gain little advantage from
adding these semantics. There are advantages for the reader, but in this case
not for both author and reader. As a result, we will probably add such support
to kcite; but, if we are honest, find it unlikely that when acting as content
authors, we will find the time to add this additional semantics.

Second, our presentation of mathematics could be modified to automatically
generate MathML from any included \TeX{} markup. The transformation could be
performed on the server, using MathJax; MathML would still be rendered on the
client to webfonts. This would mean that any embedded maths would be
discoverable because of the existence of MathML, which is a considerable
advantage. However, neither the reader nor the author gain any advantage from
doing this, while paying the cost of the slower load times and higher server
load that would result from running JavaScript on the server. More over, they
would pay this cost regardless of whether their content were actually being
consumed computationally. As the situation now stands, the computational user
needs to identify the insert of MathJax into the web page, and then transform
the page using this library, none of which is standard. This is clearly a
serious compromise, but we feel a necessary one.

Our support for microarrays offers the possibility of the most specific and
increased level of semantics of all of our plugins. Knowledge about a species
or a microarray experimental design can be precisely represented.
However, almost by definition, this form of knowledge is fairly niche and only
likely to be of relevance to a small community. However, we do note that the
knowledgeblog process based around commodity technology does offer a
publishing process that can be adapted, extended and specialised in this way
relatively easily. Ultimately the many small communities that make up the
long-tail of scientific publishing adds up to one large one.

\section{Conclusion}

Semantic publishing is a desirable goal, but goals need to be realistic and
achievable. to move towards semantic publishing in \kblog, we have tried to
put in place an approach that gives benefit to readers, authors and
computational interpretation. As a result, at this stage, we have light
semantic publishing, but with small, but definite benefits for all.

Semantics give meaning to entities. In \kblog, we have sought benefit by
``saying'' within the \kblog environment that entity \emph{x} is either
\textbf{maths}, a \textbf{citation} or a \textbf{microarray} data entity
reference. This is sufficient for the \kblog infra-structure to ``know what to
do'' with the entity in question. Knowing that some publishable entity is a
``lump'' of maths tells the infra-structure how to handle that entity: the reader
has benefit from it looking like maths; the author has benefit by not having
to do very much; and the infra-structure knows what to do. In addition, this
approach leaves in hooks for doing more later.

It is not necessarily easy to find compelling examples that give advantages
for all steps. Adding in CiTO attributes to citations, for instance, has
obvious advantages for the reader, but not the author. However, advantages may
be indirect; richer reader semantics may give more readers and thus more
citations---the thing authors appreciate as much as the act of publishing
itself. It is, however, difficult to imagine how such advantages can be
conveyed to the author at the point of writing. It is easy to see the
advantages of semantic publishing for readers, as a community we need to pay
attention to advantages to the authors. Without these ``carrots'', we will only
have ``sticks'' and authors, particularly technically skilled ones, are highly
adept at working around sticks.

\bibliography{3_steps}

\begin{thebibliography}{10}

\bibitem{shadbolt2006semantic}
Shadbolt, N., Hall, W., Berners-Lee, T.:
\newblock The semantic web revisited.
\newblock Intelligent Systems, IEEE \textbf{21}(3) (2006)  96--101

\bibitem{shotton2009semantic}
Shotton, D.:
\newblock Semantic publishing: the coming revolution in scientific journal
  publishing.
\newblock Learned Publishing \textbf{22}(2) (2009)  85--94

\bibitem{shotton2009adventures}
Shotton, D., Portwin, K., Klyne, G., Miles, A.:
\newblock Adventures in semantic publishing: exemplar semantic enhancements of
  a research article.
\newblock PLoS computational biology \textbf{5}(4) (2009)  e1000361

\bibitem{bizer2009linked}
Bizer, C., Heath, T., Berners-Lee, T.:
\newblock Linked data-the story so far.
\newblock International Journal on Semantic Web and Information Systems
  (IJSWIS) \textbf{5}(3) (2009)  1--22

\bibitem{latex}
Landport, L.:
\newblock The \LaTeX{} book.
\newblock Adison wesley, Reading, MA (1984)

\bibitem{stlr_kblog_2011}
Lord, P., Cockell, S., Swan, D.C., Stevens, R.:
\newblock The ontogenesis knowledgeblog: Lightweight publishing about
  semantics, with lightweight semantic publishing.
\newblock In: Semantic Web Technologies for Libraries and Readers. (2011)

\bibitem{wordpress}
Wordpress:
\newblock \url{http://www.wordpress.org}.

\bibitem{Hull:2006:Nucleic-Acids-Res:16845108}
Hull, D., Wolstencroft, K., Stevens, R., Goble, C., Pocock, M.R., Li, P., Oinn,
  T.:
\newblock Taverna: a tool for building and running workflows of services.
\newblock Nucleic Acids Res \textbf{34}(Web Server issue) (Jul 2006)  729--732

\bibitem{tex}
Knuth, D.E.:
\newblock The \TeX{} Book. 3rd edition edn.
\newblock Adison Wesley, Reading, MA (1986)

\bibitem{wp-latex}
{WP Latex}:
\newblock \url{http://wordpress.org/extend/plugins/wp-latex/}.

\bibitem{mathjax-latex}
Mathjax-Latex:
\newblock \url{http://wordpress.org/extend/plugins/mathjax-latex/}.

\bibitem{mathjax}
Mathjax:
\newblock \url{http://www.mathjax.org}.

\bibitem{Mendeley}
Mendeley:
\newblock \url{http://www.mendeley.org}.

\bibitem{zotero}
Zotero:
\newblock \url{http://www.zotero.org}.

\bibitem{Brazma01012003}
Brazma, A., Parkinson, H., Sarkans, U., Shojatalab, M., Vilo, J.,
  Abeygunawardena, N., Holloway, E., Kapushesky, M., Kemmeren, P., Lara, G.G.,
  Oezcimen, A., Rocca-Serra, P., Sansone, S.A.:
\newblock {ArrayExpress- public repository for microarray gene expression data
  at the EBI}.
\newblock Nucleic Acids Research \textbf{31}(1) (2003)  68--71

\bibitem{cito}
Shotton, D.:
\newblock {CiTO, the Citation Typing Ontology}.
\newblock Journal of Biomedical Semantics \textbf{1}(Suppl 1) (2010) ~S6

\end{thebibliography}
\label{sec:references}
\bibliographystyle{splncs}

\end{document}